\documentclass[]{aa}
\usepackage{psfig}
\usepackage{txfonts}
\usepackage{graphicx}
\usepackage{amssymb}
\usepackage[below]{placeins}
\usepackage{natbib}
\bibpunct{(}{)}{;}{a}{}{,}

\def \arcmin{$^{\prime}$}

\begin{document}

\title{Possible non-thermal nature of the soft-excess emission in the cluster of galaxies S\'ersic~159-03}
\author{N. Werner\inst{1}
 \and J.S. Kaastra\inst{1,2}
 \and Y. Takei\inst{1,3}
 \and R. Lieu\inst{4}
 \and J. Vink\inst{2}
 \and T. Tamura\inst{3}}

\offprints{N. Werner, email {\tt n.werner@sron.nl}}

\institute{     SRON Netherlands Institute for Space Research, Sorbonnelaan 2,
                NL - 3584 CA Utrecht, the Netherlands
\and		Astronomical Institute, Utrecht University, P.O. Box 80000,
                NL - 3508 TA Utrecht, the Netherlands 
\and		Institute of Space and Astronautical Science, JAXA, 3-1-1 Yoshinodai, Sagamihara, Kanagawa 229-8510, Japan
\and		Department of Physics, University of Alabama, Huntsville, AL 35899, USA                
}

\date{Received, accepted }

\abstract{We present an analysis of new Suzaku data and archival data from XMM-Newton of the cluster of galaxies S\'ersic 159-03, which has a strong soft X-ray excess emission component. The Suzaku observation confirms the presence of the soft excess emission, but it does not confirm the presence of redshifted \ion{O}{vii} lines in the cluster. Radial profiles and 2D maps derived from XMM-Newton observations show that the soft excess emission has a strong peak at the position of the central cD galaxy and the maps do not show any significant azimuthal variations. Although the soft excess emission can be fitted equally well with both thermal and non-thermal models, its spatial distribution is neither consistent with the models of intercluster warm-hot filaments, nor with models of clumpy warm intracluster gas associated with infalling groups. Using the data obtained by the XMM-Newton Reflection Grating Spectrometers we do not confirm the presence of the warm gas in the cluster centre with the expected properties assuming the soft excess is of thermal origin. 
The observed properties of the soft excess emission are consistent with the non-thermal interpretation. While the high density of relativistic electrons associated with the peak of the soft emission in the cluster centre might have been provided by an active galactic nucleus in the central cD galaxy, the underlying population might have been accelerated in diffuse shocks. 

}  

\maketitle

\section{Introduction}
\label{intro}

In the past 10 years, extreme ultraviolet (EUV) and soft X-ray observations (with {\it{EUVE}}, {\it{ROSAT}}, {\it{BeppoSAX}}, {\it{XMM-Newton}}) unveiled emission in excess of that expected from the thermal intracluster medium (ICM) in a number of clusters of galaxies. The first papers reporting the discovery of the excess EUV emission from clusters interpreted their finding as thermal emission from warm diffuse gas \citep{lieu1996a,lieu1996b,mittaz1998}. However, the large mass of the warm gas required to explain the soft excess emission, which would cool very rapidly at the derived temperatures, was a problem for the model \citep{mittaz1998}. 
As an alternative explanation for the cluster soft excess inverse Compton (IC) emission by cosmic-ray electrons scattering off the cosmic microwave background (CMB) was proposed \citep{hwang1997,ensslin1998,sarazin1998,lieu1999}. \citet{kaastra2003} reported the discovery of soft excess and \ion{O}{vii} line emission around five clusters observed with XMM-Newton. They attributed this component to emission from intercluster filaments of the Warm-Hot Intergalactic Medium in the vicinity of the clusters. \citet{nevalainen2006} reanalized the XMM-Newton data of 4 clusters with a reported soft excess emission using the newest calibration and found that the XMM-Newton EPIC instruments differ on the magnitude of the soft excess in all clusters. While EPIC/MOS data still show soft excess emission in all clusters reported by \citet{kaastra2003}, EPIC/pn data show the soft excess emission only in the cluster S\'ersic~159-03 and the redshifted \ion{O}{vii} line emission is not observed in the reprocessed data. \citet{nevalainen2006} conclude that the possibility that the reported \ion{O}{vii} line emission in the outskirts of these clusters is due to heliospheric or geocoronal charge exchange emission, as suggested by \citet{bregman2006}, cannot be ruled out. 

The rich southern cluster of galaxies S\'ersic~159-03, also known 
as \object{ACO~S1101}, was discovered by \citet{sersic1974}. It is a nearby \citep[z=0.0564;][]{maia1987} X-ray bright cluster with a luminosity of $L_{\mathrm{X}}=5.35\times10^{44}$~erg~s$^{-1}$ in the 0.5--2.0 keV band \citep{degrandi1999}. The cluster is relaxed with no obvious peculiarities in the X-ray morphology \citep{kaastra2001}. The temperature profile peaks at k$T=2.7$~keV at a radius of $\sim2.5$\arcmin\ from the core \citep{deplaa2006}. The temperature drop in the core is relatively modest and the temperature outside the radius of 4\arcmin\ drops rapidly by at least a factor of $\sim2$ \citep{kaastra2001,deplaa2006}. 
S\'ersic~159-03 has the strongest soft excess emission of all clusters observed with XMM-Newton, with the detection well above the calibration uncertainty. Its soft excess was discovered using ROSAT PSPC data by \citet{bonamente2001} and confirmed by two independent XMM-Newton observations \citep{kaastra2003,bonamente2005,nevalainen2006}. Analysing a deep (122~ks) XMM-Newton observation, \citet{bonamente2005} show that strong soft excess emission is detected out to the radial distance of 12\arcmin\ (0.9~Mpc). They interpret their data using two models: by invoking a warm reservoir of thermal gas, and by relativistic electrons that are part of a cosmic ray population. In their conclusions they slightly favor the thermal interpretation. Based on the analysis of the same data set, \citet{deplaa2006} propose as a potential source of the soft excess non-thermal emission arising from IC scattering between CMB photons and relativistic electrons, that are accelerated in bow shocks associated with ram pressure stripping of in-falling galaxies. 

Here we analyse the soft excess in S\'ersic 159-03 using a new deep Suzaku \citep{mitsuda2007} observation, together with two archival XMM-Newton \citep{jansen2001} observations. We analyse the Suzaku XIS1 \citep{koyama2007} data in order to obtain a confirmation of the soft excess with an independent instrument. We compare the observed soft excess flux in the three analysed observations and determine the systematic uncertainty on its value. Furthermore, Suzaku has a superior spectral redistribution function at low energies, which allows us to resolve the oxygen line emission with unprecedented accuracy. The detection of \ion{O}{vii} line emission with Suzaku would prove the thermal origin of the soft excess emission and determining its redshift would allow us to confirm its cluster origin. We also take advantage of the excellent statistics of the archival XMM-Newton observations, which allow us to analyse the spatial distribution of the soft excess emission in the cluster. 

The paper is organised in the following way. In section 2, we describe the Suzaku and XMM-Newton data of S\'ersic~159-03 and discuss the spectral modeling. In section 3, we fit the soft excess emission detected by Suzaku with thermal and non-thermal models, we compare the soft excess flux detected by Suzaku with the flux detected during two XMM-Newton observations. Furthermore, we exploit the higher signal-to-noise and good spatial resolution of the XMM-Newton observations to investigate the spatial structure of the soft excess emission by analysing radial profiles and 2D maps, and we search for possible lines from warm gas in the cluster core with the Reflection Grating Spectrometers (RGS) on XMM-Newton. In section 4, we discuss the physical interpretation of the results and finally, in section 5, we summarise the main conclusions. In the Appendix we describe the details of the calibration of the contaminating layer on the optical blocking filter of the XIS1 detector on Suzaku.

Throughout the paper we use $H_{0}=70$ km$\, $s$^{-1}\, $Mpc$^{-1}$, $\Omega_{M}=0.3$, $\Omega_{\Lambda}=0.7$, which imply a linear scale of 73~kpc\, arcmin$^{-1}$ at the cluster redshift of $z=0.0564$ \citep{maia1987}. 
Unless specified otherwise, all errors are at the 68\% confidence level for one interesting parameter ($\Delta \chi^{2}=1$). Upper limits are at the $2\sigma$ confidence level. The elemental abundances are given with respect to the proto-solar values of \citet{lodders2003}.

\section{Observations and data analysis}
\subsection{Suzaku data}
\label{suzakudata}

\begin{figure}
\includegraphics[width=6.7cm,clip=t,angle=0.]{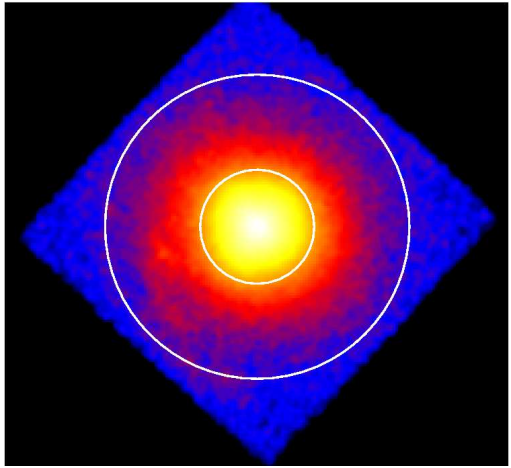}
\caption{Suzaku XIS image of S\'ersic~159-03 in the $1$--$4$~keV band. The data obtained by the four XIS detectors were combined. The white circles indicate the extraction regions for the spectral analyses: circular region with a radius of 3\arcmin\ (the smaller circle), and annulus with inner and outer radii of 3\arcmin\ and 8\arcmin, respectively.}
\label{image}
\end{figure}

S\'ersic~159-03 was observed with Suzaku between April 26--28, 2006 with a total exposure time of 64~ks. 

We analyse the data obtained by the back illuminated XIS1 detector, using event files of version 1.2 products. We only use the XIS1 data and not the remaining 3 front illuminated XIS detectors (0, 2, 3) in order to reduce the systematic uncertainties. The XIS1 detector has the largest effective area and is best calibrated at the low energies of all XIS instruments. XIS1 has also the best studied radial profile of the contamination of the optical blocking filter.

The XIS CCD camera covers the 0.2--12 keV energy range with an energy resolution of 40~eV at 0.5~keV and 130~eV at 5.9~keV. It has a square field of view of 18\arcmin$\times$18\arcmin. Its advantage is the absence of a low energy tail in the pulse-height distribution function, which makes it an excellent instrument to study line emission at low energies. 

The image of S\'ersic~159-03 in the 1--4~keV band is shown in Fig.~\ref{image}. Vignetting effects were not taken into account for this image and the background events were not subtracted. Suzaku XIS1 has a relatively broad point spread function with a half-power diameter of 2.3\arcmin\ \citep{serlemitsos2007}. Therefore, the extraction regions also have to be correspondingly large. The circles indicate two regions, that we used for spectral analysis. The spectra were extracted from a circular region with a radius of 3\arcmin\ centred on the cluster and from an annulus with inner and outer radii of 3\arcmin\ and 8\arcmin, respectively. We extracted the spectra using XSELECT distributed with HEASOFT 6.1.1. 

To subtract the instrumental background, we extracted spectra from night Earth observations. The level of the instrumental background is anti-correlated with the cut-off rigidity (COR) value at the position of the satellite. We extracted night Earth spectra for each 1 GeV~$c^{-1}$ interval of COR and added them weighted by exposure time for the given COR range in the source observation. The background spectra were extracted from the same detector region as the source spectra in order to avoid possible systematic effects due to spatial variation of the instrumental background. Cosmic background components are included in the spectral fitting phase (see Sect. \ref{sectback}).  

The spectral redistribution files (RMF) were created using `xisrmfgen'. 
The ancillary response files (ARF) were created using `xissimarfgen', which is based on ray-tracing \citep{ishisaki2006}. We created a separate ARF for both extraction regions. For the incident flux distribution, we adopted the XIS image of the cluster in the $1.0$--$4.0$ keV band.  The effective area of the XIS detectors below 1~keV is affected by carbon and oxygen contamination of the optical blocking filter. The contamination is time and position dependent. 
To model the column and the position dependence of the contaminant, we use the  "ae\_xi1\_contami\_20061016.fits" contamination table, which contains the radial profiles and the best estimated values of the contaminating column for different time periods. For the precise value of the contaminating column density in the time of our observation, we use the values determined using simultaneous calibration observations of the blazar PKS~2155-304 performed by Chandra LETGS/HRC, XMM-Newton and Suzaku on May 1st 2006, only 3 days after our observation (see the Appendix for the details of the analysis). 
We found that the column densities of the contamination that are suggested by the XIS team ($N_{\mathrm{C}}=3.5\times10^{18}$~cm$^{-2}$, $N_{\mathrm{O}}=5.9\times10^{17}$ cm$^{-2}$) are not consistent with the calibration observation of PKS~2155-304.
The carbon and oxygen contaminating columns determined using the simultaneous observation of Suzaku and Chandra LETGS/HRC are $N_{\mathrm{C}}=(3.78\pm0.05)\times10^{18}$~cm$^{-2}$ and $N_{\mathrm{O}}=6.30\pm0.08\times10^{17}$~cm$^{-2}$, for the centre of the field of view. 
Since the most probable composition of the contaminant is C$_{24}$H$_{38}$O$_{4}$, the C/O ratio was fixed in the fitting process to 6 (see Appendix for the details). The contaminating columns determined using the simultaneous Suzaku and XMM-Newton EPIC/pn observation are $N_{\mathrm{C}}=(4.11\pm0.06)\times10^{18}$~cm$^{-2}$ and $N_{\mathrm{O}}=(6.2\pm0.5)\times10^{17}$~cm$^{-2}$, for the centre of the field of view. This analysis shows, that there is $\sim$10\% uncertainty in the value of the contaminating column. 

In order to provide a more conservative estimate for the soft excess, we will adopt in this paper the values for the contaminating column in the centre of the field of view determined by the simultaneous observation of Suzaku and Chandra LETGS/HRC: $N_{\mathrm{C}}=3.78\times10^{18}$~cm$^{-2}$ and $N_{\mathrm{O}}=6.30\times10^{17}$~cm$^{-2}$. These values correspond to the nominal values in the standard contamination file for July 1st 2006. To illustrate how sensitive the best fit soft excess flux is to the precise value of the contamination, we will also compare our results in Sect.~\ref{excess} with those obtained assuming the nominal column densities in the contamination file for the date when our observation was performed ($N_{\mathrm{C}}=3.5\times10^{18}$~cm$^{-2}$) and for September 1st 2006 ($N_{\mathrm{C}}=4.2\times10^{18}$~cm$^{-2}$). We note that while the former value is too low and is not consistent with the values obtained from the analysis of the cross-calibration data, the second value is only slightly higher than that obtained by the cross-calibration of EPIC/pn and XIS1.

We inspected the solar proton flux measured with the ACE and Wind satellites during our observation. The proton flux increases in the last 24~ks of the observation to $\sim$$10^{9}$~s$^{-1}$~cm$^{-2}$, the level at which the XIS1 spectrum is affected in \citet{fujimoto2006}. In order to investigate the possibility of the contamination of our data by charge exchange emission in the Earth's magnetosphere, we extracted a spectrum using only the first 40 ks, with a low proton flux. We compared it with the spectrum extracted from the remaining 24 ks of the Suzaku observation. We verified that the spectrum did not change and that the elevated soft proton level in the second part of the observation did not cause a detectable contamination by charge exchange emission. Hence we used the whole observation in our analyses. 

\subsection{XMM-Newton data}

S\'ersic~159-03 was observed with XMM-Newton on May 5, 2000 and on November 20--21, 2002 with a total exposure time of 60~ks and 122~ks, respectively. The data were reduced with the 7.0.0 version of the XMM Science Analysis System (SAS). The data processing and the subtraction of the instrumental background for the European Photon Imaging Cameras (EPIC) was performed as described in \citet{deplaa2006}. Cosmic background components in the EPIC data are included in the spectral fitting phase. The RGS data were processed as described in \citet{deplaa2006}. 

\subsection{Spectral analysis}
\label{spex}
For the spectral analysis we use the SPEX package \citep{kaastra1996}. 
We fix the Galactic absorption in our model to the value deduced from \ion{H}{i} data $N_{\mathrm{H}}=1.79\times10^{20}$~cm$^{-2}$ \citep{dickey1990}. We use the \citet{verner1996} cross-sections in our model of Galactic absorption. 
For the spectral fitting of the cluster spectra, we use the {\it{wdem}} model \citep{kaastra2004}, which proved to be the most successful in fitting cluster cores \citep[e.g.][]{kaastra2004,deplaa2005,werner2006,deplaa2006}. This model is a differential emission measure (DEM) model with a cut-off power-law distribution of emission measures versus temperature. In our analysis, we fit both a thermal plasma model (MEKAL) and a non-thermal power-law model to the soft excess emission. In our spectral fits both the cluster and the soft excess emission are free parameters.

Unless specified otherwise, we use the 0.4--7 keV energy range for the spectral analysis of the Suzaku XIS1 data and the 0.4--10 keV band for the XMM-Newton data. The Suzaku and the two XMM-Newton data sets are fitted separately.

\subsection{Modeling of the X-ray background emission}
\label{sectback}

\begin{table}
\begin{center}
\caption{The unabsorbed soft Galactic foreground flux in units of $10^{-12}$ erg s$^{-1}$ cm$^{-2}$ deg$^{-2}$ in the $0.3$--$10$ keV band. The column RASS shows the fluxes derived from the parameters reported by \citet{kuntz2000}. In columns Suzaku A and B, we report the values derived from two offset pointings near A2218 by \citet{takei2006}. The values in column S159-03 were derived using XMM-Newton data in the $9$\arcmin--$12$\arcmin\ annulus around the cluster S\'ersic 159-03 by \citet{deplaa2006}.
\label{tab:back}}
\begin{tabular}{lcccc}
\hline
\hline
Component		   &	RASS 	& 	Suzaku~A 	& Suzaku~B 	& S159-03   \\
\hline
LHB k$T$	   	   & 0.082      &        0.08	        &  0.08    	&   0.07    \\
LHB flux                   & 1.22       &	 1.74		&  3.35		&   2.23    \\
SDC k$T$ 		   & 0.068      &         --   		&  --  	   	&   --      \\
SDC flux 		   & 0.87       &         --		&  --		&   -- 	    \\
HDC k$T$ 		   & 0.127      &        0.16  	        &  0.25    	&   0.20    \\  
HDC flux 		   & 2.44       &	 1.85		&  2.06		&   12.20   \\
total soft flux            & 4.53       &        3.59           &  5.41         &   14.43   \\

\hline
\end{tabular}
\label{tab:cxb}
\end{center}
\end{table}

We correct for the Cosmic X-ray Background (CXB) during spectral fitting.
\citet{kuntz2000} distinguish 4 different background/foreground components: the extragalactic power-law (EPL), the local hot bubble (LHB), the soft distant component (SDC) and the hard distant component (HDC). The EPL component is the integrated emission of faint discrete sources, mainly distant Active Galactic Nuclei (AGNs).
The LHB is a local supernova remnant, in which our Solar System resides. It produces virtually unabsorbed emission at a temperature of $\sim$10$^{6}$~K. The soft and hard distant components originate at larger distances. They might be identified with the Galactic halo, Galactic corona or the Local group emission and are absorbed by almost the full Galactic column density. In the first column of Table~\ref{tab:cxb} we show the fluxes and temperatures for the Galactic foreground derived from the parameters reported by \citet{kuntz2000}, based on the Rosat All Sky Survey data. Using the spectral band above 0.4 keV we can not reliably distinguish the SDC emission from the LHB component. Therefore, at temperatures below 0.1~keV we  only consider the contribution of the LHB. 

Using data obtained by XMM-Newton, \citet{deluca2004} found that the photon index of the EPL is $\Gamma=1.41\pm0.06$ and its 2--10~keV flux is $(2.24\pm0.16)\times10^{-11}$~erg~cm$^{-2}$~s$^{-1}$~deg$^{-2}$ (90\% confidence level). By fitting the spectra extracted from the 9\arcmin--12\arcmin\ region around S\'ersic~159-03 and assuming a photon index of $\Gamma=1.41$, \citet{deplaa2006} found a $2$--$10$~keV EPL flux of $2.26\times10^{-11}$~erg~cm$^{-2}$~s$^{-1}$~deg$^{-2}$ (corresponding to a 0.3--10.0 keV flux of $3.14\times10^{-11}$~ergs~cm$^{-2}$~s$^{-1}$~deg$^{-2}$). 
We adopt these values in our model of the EPL background. 
While the measured mean variation in the EPL emission in the $2$--$10$~keV band in the sky is only $\sim3.5$\%, the intensity of the soft emission in the $0.2$--$1.0$ keV band varies by $\sim35$\% from field to field \citep{lumb2002,kaastra2003}. Unfortunately, the emission from S\'ersic~159-03 fills the entire field of view of both Suzaku and XMM-Newton, which does not allow us to determine the properties of the local background directly from an annulus in the outer part of the field of view of the instruments. As shown in Table~\ref{tab:cxb}, \citet{deplaa2006} found, by fitting the spectra extracted from the 9\arcmin--12\arcmin\ region around the cluster, a flux of the HDC emission which was more than 5 times higher than the HDC flux in the soft band measured by ROSAT \citep{kuntz2000}, and determined from two Suzaku blank field observations \citep{takei2006}. Since the flux at low energies in the outer parts of the cluster is not consistent with the values determined in other fields, large part of it may still be associated with the soft excess emission, which we are looking for. 

We inspected the ROSAT all sky survey data in the 3/4 keV band (0.5--0.9~keV; which is less affected by the variations in the Galactic absorption) and in the 1/4 keV band (0.1--0.4~keV) around S\'ersic~159-03 and found that the variations in the soft emission around the cluster are small and there is no indication of a strongly elevated level of Galactic foreground. The ROSAT count rates in the vicinity of S\'ersic~159-03 are consistent with those in the Suzaku offset pointing B at the cluster Abell~2218 \citep{takei2006}. Therefore, for the soft foreground emission, we adopt the fluxes reported for the offset pointing B by \citet{takei2006}. The sum of the adopted fluxes of the soft foreground components is $\sim20$\% higher than the average soft X-ray background flux determined by \citet{kuntz2000} from the ROSAT observations.

\section{Results}

\begin{figure*}[tb]
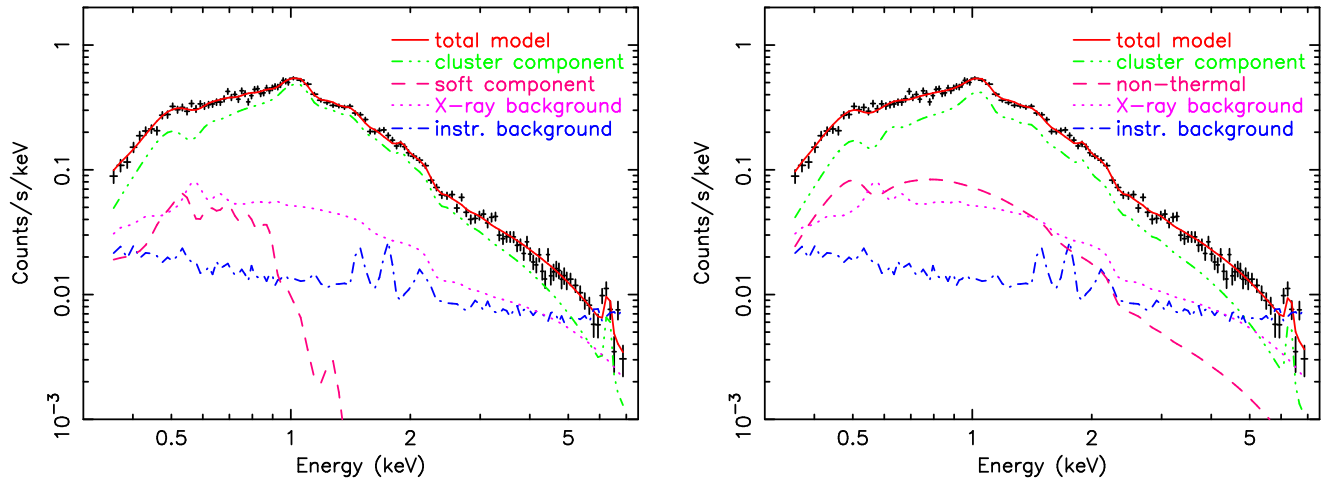

\begin{minipage}{0.5\textwidth}
\includegraphics[width=0.7\textwidth,clip=t,angle=270.]{softmodel.ps}
\end{minipage}
\begin{minipage}{0.5\textwidth}
\includegraphics[width=0.7\textwidth,clip=t,angle=270.]{nontherm.ps}
\end{minipage}\\
\caption{{\it Left panel}: Suzaku XIS1 spectrum of the $3$\arcmin--$8$\arcmin\ annulus around the cluster. The total model and its components are plotted separately. The soft excess emission is fitted with a thermal model. The dash-dotted line indicates the subtracted instrumental background, the dotted line indicates the cosmic X-ray background, the dashed line shows the thermal emission associated with the soft excess, the dash-dotted-dotted-dotted line indicates the cluster emission and the full line indicates the total model. {\it Right panel}: The same spectrum as on the left panel, with the soft excess modelled as non-thermal emission. } 
\label{XIS1}
\end{figure*}

\begin{table}
\caption{Fit results for the Suzaku XIS1 data of S\'ersic~159-03. We fit the multi-temperature $wdem$ model to the cluster emission, combined with both thermal and non-thermal models to describe the soft excess in a circular region with a radius of 3\arcmin\ centred on the cluster core and in an annulus with inner and outer radii of 3\arcmin\ and 8\arcmin, respectively. For the Galactic absorption we use a value of $1.79\times10^{20}$~cm$^{-2}$. Emission measures ($Y = \int n_{\mathrm{e}} n_{\mathrm{H}} dV$) are given in 10$^{66}$ cm$^{-3}$ and the power-law normalisations ($N_{\mathrm{NT}}$) are given in $10^{51}$ photons~s$^{-1}$~keV$^{-1}$. The absorbed fluxes ($F$) were determined in the 0.3--10.0 keV band using an $N_{\mathrm{H}}$ value of $1.79\times10^{20}$~cm$^{-2}$, and they are given in $10^{-10}$~ergs~s$^{-1}$~cm$^{-2}$~deg$^{-2}$. Abundances are given with respect to the proto-solar values of \citet{lodders2003}.}
\begin{center}
\begin{tabular}{l|cccc}
\hline\hline
Par.		 & 0\arcmin--3\arcmin\  & 0\arcmin--3\arcmin\ & 3\arcmin--8\arcmin\ & 3\arcmin--8\arcmin\   \\
		 & therm.	 	    & non-therm.      &   therm. 	  & non-therm.     \\
\hline
$Y_{\mathrm{cl}}$	 &  $15.8\pm0.2$    &  $14.9\pm0.5 $  &  $3.71\pm0.09$    &  $3.2\pm0.3$   \\
$F_{\mathrm{cl}}$	 &  $30.68\pm0.38$    &  $29.03\pm1.02$   &	 $1.03\pm0.03$    &   $0.92\pm0.06$  \\
$kT_{\mathrm{mean}}$     &  $2.61\pm0.08$   &  $2.56\pm0.09$  &  $2.41\pm0.14$    &  $2.37\pm0.19$  \\
O                        &  $0.46\pm0.12$   &  $0.64\pm0.08$  &  $<0.30$	  &  $0.51\pm0.15$  \\
Si                       &  $0.48\pm0.04$   &  $0.50\pm0.04$  &  $0.21\pm0.07$    &  $0.25\pm0.08$  \\
S			 &  $0.38\pm0.05$   &  $0.41\pm0.06$  &  $<0.52$          &  $0.15\pm0.11$  \\
Ar			 &  $0.31\pm0.13$   &  $0.35\pm0.14$  &	 $<0.32$	  &  $<0.36$   	\\
Ca			 &  $0.72\pm0.18$   &  $0.78\pm0.19$  &  $<0.62$          &  $<0.46$  \\
Fe			 &  $0.58\pm0.02$   &  $0.60\pm0.02$  &  $0.39\pm0.03$    &  $0.41\pm0.04$  \\
$Y_{\mathrm{exc}}$	 &  $0.6\pm0.2$     & 	    --	      &  $1.25\pm0.25$    &       --      \\
$kT$			 &  $0.31\pm0.05$   &       --        &  $0.19\pm0.01$    &       --      \\
O			 &  0.3		    &       --	      &  $0.11\pm0.04$	  &       --	  \\
$N_{\mathrm{NT}}$	 &	--	    &	 $3\pm1$      &       --          & $2.0\pm0.7$   \\
$\Gamma$		 &      --          &$2.11^{+0.27}_{-0.15}$ &   --        &  $2.41^{+0.33}_{-0.19}$  \\
$F_{\mathrm{exc}}$	 &  $0.48\pm0.13$       &  $2.42\pm0.76$ &  $0.08\pm0.01$      &  $0.23\pm0.08$      \\
\hline
$\chi^2$ / $\nu$	 &  203/128    	    &  212/128        &  154/108          & 164/108	 \\
\hline
\end{tabular}
\label{tabprof}
\end{center}
\end{table}

\subsection{Thermal and non-thermal models of soft excess emission}
\label{xis1spectr}

We analyse the Suzaku XIS1 spectra extracted from the two regions shown in Fig.~1: a circular region with a radius of 3\arcmin, and annulus with inner and outer radii of 3\arcmin\ and 8\arcmin, respectively. We fit the cluster emission with the multitemperature $wdem$ model (see Sect. \ref{spex}). The excess emission at low energies can be formally fitted as strongly subgalactic absorption column density ($N_{\mathrm{H}}$). When fitting the cluster spectra with free $N_{\mathrm{H}}$, we obtain for the inner extraction region an absorption column density of $N_{\mathrm{H}}=(1.15\pm0.20)\times10^{20}$~cm$^{-2}$ and for the outer region we obtain a $2\sigma$ upper limit of $N_{\mathrm{H}}<1.4\times10^{19}$~cm$^{-2}$. Both values are well below the Galactic value toward the cluster $N_{\mathrm{H}}=1.79\times10^{20}$~cm$^{-2}$ \citep[for an extensive discussion on absorption toward S\'ersic~159-03 see][]{bonamente2001}. In the rest of the analysis, we fix the $N_{\mathrm{H}}$ in our spectral model to the Galactic value.

In order to describe the soft excess emission, we try to fit both thermal and non-thermal models. Both the parameters of the cluster component (ICM emission measure, temperature structure, and abundances) and of the soft excess emission are free to vary in the fitting process. When fitting the soft excess emission with the thermal model, we fix the abundances of the warm-hot gas other than oxygen to 0.3 solar and fit its emission measure, temperature and oxygen abundance. 
In the non-thermal model, the power-law emission caused by relativistic electrons is assumed to contribute both to the soft excess and some part of the higher energy emission.
When fitting the soft excess emission with a non-thermal component, we fit the photon index and the normalisation of the power-law. 

In Table~\ref{tabprof} and in Fig.~\ref{XIS1}, we show the best fit results using both thermal and non-thermal models to fit the soft excess. We also show the best fit results for the cluster emission, including the best fit abundance values for 6 elements. When fitting the data from the inner extraction region the oxygen abundance of the soft excess component is not constrained by the data and we fix its value to 0.3 solar together with the other abundances. The best fit temperature of the warm component does not depend on the adopted abundance values. For lower adopted metallicity, the best fit emission measure of the warm gas will increase and for higher metallicities it will decrease. 
The soft excess flux in the cluster centre is higher than in the outer extraction region, which indicates that the soft excess peaks at the cluster core. 

As we show in Table~\ref{tabprof} and in Fig.~\ref{XIS1}, thermal and non-thermal models fit the data statistically equally well. We note that the reduced $\chi^2$ values of our best fit models are relatively high because we did not include systematic errors on the data in the spectral fitting process and at the high number of counts of our dataset the systematic errors dominate over the statistical errors.
Clear detection of a redshifted \ion{O}{vii} line emission would be an unambiguous proof for the thermal origin of the soft excess emission. 
Since XIS1 has a much better spectral redistribution function at low energies than EPIC, we use its data for the outer extraction region to search for the redshifted \ion{O}{vii} line emission.
In order to model the broad soft excess we fit a power-law, and subsequently add a narrow Gaussian at the expected energy of the \ion{O}{vii} intercombination line at the cluster redshift. We find that the power-law describes the soft excess emission sufficiently well and there is no need for the additional Gaussian. This way, we find a formal upper limit on the \ion{O}{vii} line emission from the cluster of $1.69\times10^{-7}$~photons~cm$^{-2}$~s$^{-1}$~arcmin$^{-2}$, which is comparable to that found by \citet{takei2006} for the cluster Abell~2218. This demonstrates the sensitivity of XIS for lines at low energies. Although the thermal model of the excess emission which already contains the \ion{O}{vii} line emission (with a line flux of ($6.4\pm1$)$\times10^{-7}$~photons~cm$^{-2}$~s$^{-1}$~arcmin$^{-2}$) is consistent with the data, 
we can not confirm the presence of the redshifted \ion{O}{vii} line emission which would be the clear proof for the thermal origin of the soft excess. In subsection~\ref{RGSsubsec} we show that the \ion{O}{vii} line emission is also not present in the RGS data, as would be expected given the spatial distribution of the excess emission (see Sect.~\ref{prof}).

\subsection{Comparison with XMM-Newton soft excess detections and the systematic uncertainties}
\label{excess}

\begin{figure}
\includegraphics[width=6.7cm,clip=t,angle=270.]{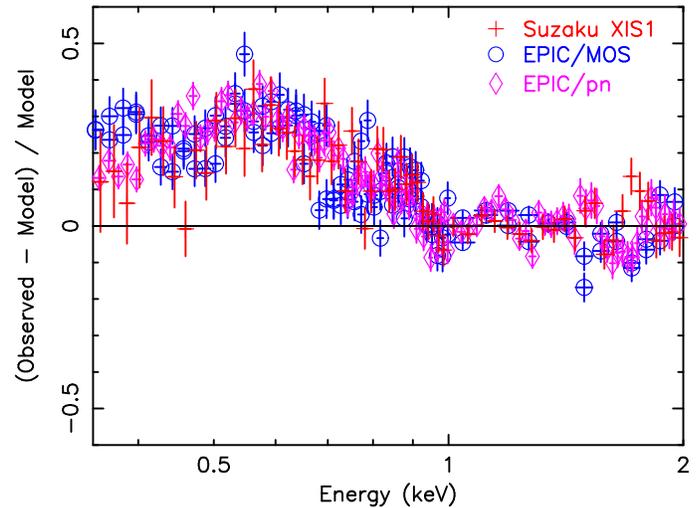}
\caption{The residuals of the best fit cluster model in the 0.9--7.0 keV band to the spectra obtained by Suzaku XIS1, EPIC/MOS, and EPIC/pn from the $3$\arcmin--$8$\arcmin\ annulus around the cluster, extrapolated to low energies. All instruments show a soft excess emission below 1~keV. } 
\label{excessfig}
\end{figure}

\begin{table}
\begin{center}
\caption{The absorbed soft excess flux in units of $10^{-11}$ erg s$^{-1}$ cm$^{-2}$ deg$^{-2}$ in the $0.3$--$1.0$ keV band, determined using an $N_{\mathrm{H}}$ value of $1.79\times10^{20}$~cm$^{-2}$. The soft excess was modelled with a thermal model. We show the soft excess in two extraction regions. The datasets obtained during the first and second observation with XMM-Newton are labelled EPIC~1 and EPIC~2, respectively. For the XIS1 instrument, we report the soft excess fluxes obtained for 3 different contaminating columns. See the text for the details. 
\label{tab:back}}
\begin{tabular}{lcc}
\hline
\hline
Dataset		&	$0$\arcmin--$3$\arcmin 	&  $3$\arcmin--$8$\arcmin\  \\
EPIC~1		&	$5.95 \pm0.35$		& $0.86\pm0.08$  \\
EPIC~2		&	$6.03 \pm0.23$		& $1.12\pm0.06$  \\
XIS~A		&	$0$			& $0.23\pm0.07$  \\
XIS~B		&	$4.8\pm1.3$		& $0.75\pm0.08$ \\
XIS~C		&	$7.4\pm1.3$		& $1.29\pm0.10$  \\

\hline
\end{tabular}
\label{separ}
\end{center}
\end{table}

In order to determine the level of consistency between XMM-Newton EPIC and Suzaku XIS1 below 1~keV, we fit separately for each instrument the cluster spectra extracted from the $3$\arcmin--$8$\arcmin\ annulus in the 0.9--7.0~keV band with the multitemperature $wdem$ model (see Sect. \ref{spex}) and plot the residuals for each instrument extrapolated to low energies. 
 The residuals for Suzaku XIS1, EPIC/MOS, and EPIC/pn are shown in Fig.~\ref{excessfig}. We clearly see that all instruments consistently show a soft excess emission below 1~keV. We note that the observed excess is well above the calibration uncertainties. Observations of calibration sources with XMM-Newton do not show significant positive residuals below 1~keV.

In order to determine the flux in excess of the tail of the hot bremsstrahlung emission below 1~keV, we fit the soft excess emission with a thermal model (k$T=0.19$~keV, metallicity of 0.3 solar and emission measure as a free parameter), as a convenient empirical model. In order to provide a consistent comparison of the soft excess level between the different observations and between the two extraction regions we fit the same model to all datasets. The ICM emission measure, temperature structure and abundances are free parameters in the fitting process.
As reported by \citet{bonamente2005} and by \citet{nevalainen2006}, there is a disagreement between the XMM-Newton instruments on the level of the soft excess. We find that the spread in the best fit soft excess flux determined by fitting separately the data obtained by the individual instruments during the two XMM-Newton observations is smaller than 30\%. 
In Table~\ref{separ}, we report the intensity of the soft excess in the 0.3--1.0 keV band, obtained from individual observations by XMM-Newton and Suzaku, in the two extraction regions shown in Fig.~1. 
The two observations with XMM-Newton EPIC are indicated as ``EPIC~1'' and ``EPIC~2'', respectively. In order to indicate the possible systematic uncertainties in the results obtained with XIS1, for this instrument we report the soft excess fluxes obtained for 3 different contaminating columns: for the nominal value in the time of our observation (indicated as ``XIS~A''; $N_{\mathrm{C}}=3.5\times10^{18}$~cm$^{-2}$), for the best fit contaminating column indicated by the simultaneous calibration observation with Suzaku and Chandra LETGS/HRC (indicated as ``XIS~B''; $N_{\mathrm{C}}=3.78\times10^{18}$~cm$^{-2}$), and for the upper limit of the contaminating column indicated by the calibration observations (indicated as ``XIS~C''; $N_{\mathrm{C}}=4.2\times10^{18}$~cm$^{-2}$). We see that all observations show the soft excess emission in the 3\arcmin--8\arcmin\ annulus, and except ``XIS~A'', the 0.3--1.0 keV fluxes are $\approx1.0\times10^{-11}$ erg s$^{-1}$ deg$^{-2}$. 
Except of ``XIS~A'', the soft excess flux in the cluster centre is a factor of $\sim$6 higher than in the outer extraction region. For ``XIS~A'', we detect in the cluster core excess absorption ($N_{\mathrm{H}}=2.25\times10^{20}$ cm$^{-2}$) instead of excess emission. However, as we described in the Sect.~\ref{suzakudata} and as we show in the Appendix, the simultaneous calibration observation with Suzaku, Chandra LETGS/HRC and XMM-Newton performed only 3 days after our observation allowed us to rule out the low value for the contaminating column with which the ``XIS~A'' values were determined. The analysis of the simultaneous calibration observations shows that the contamination is between the values that were assumed in the determination of the fluxes indicated as ``XIS~B'' and ``XIS~C''. Taking into account the differences in the soft excess flux below 1~keV shown in Table~\ref{separ}, we conclude that the soft excess flux is $(6.0\pm2.0)\times10^{-11}$ erg s$^{-1}$ cm$^{-2}$ deg$^{-2}$ and  $(1.0\pm0.3)\times10^{-11}$ erg s$^{-1}$ cm$^{-2}$ deg$^{-2}$  in the inner and outer extraction region, respectively. The $\sim$20\% uncertainty in the Galactic foreground flux introduces an additional $\sim$10\% systematic uncertainty to the soft excess flux determined for the 3\arcmin--8\arcmin\ extraction region.

\subsection{Radial distribution of the soft emission}
\label{prof}

\begin{figure*}[tb]
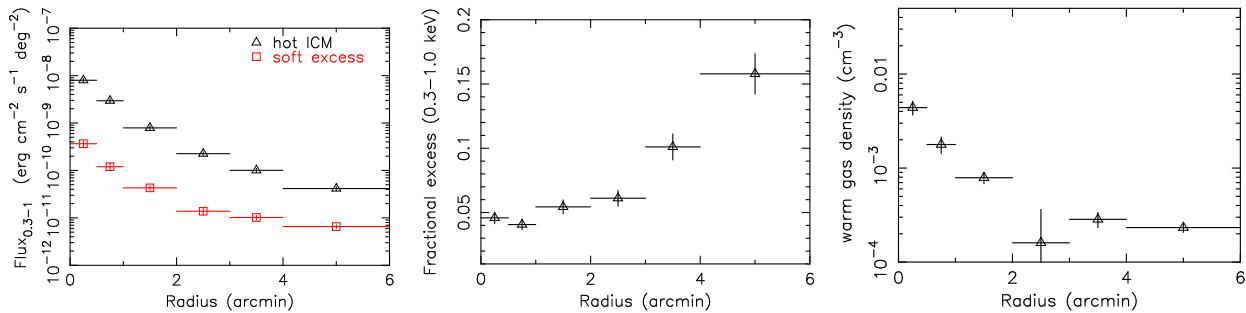

\begin{center}
\begin{minipage}{0.3\textwidth}
\includegraphics[width=0.75\textwidth,clip=t,angle=270.]{fluxes_thermlow.ps}
\end{minipage}
\begin{minipage}{0.3\textwidth}
\includegraphics[width=0.75\textwidth,clip=t,angle=270.]{fractional.ps}
\end{minipage}
\begin{minipage}{0.3\textwidth}
\includegraphics[width=0.75\textwidth,clip=t,angle=270.]{softdensity.ps}
\end{minipage}
\end{center}
\caption{{\it Left panel}: Radial profiles of the flux of the hot ICM and of the soft excess emission in the $0.3$--$1.0$ keV band derived from XMM-Newton data. The soft excess is modelled as thermal emission (k$T=0.19$~keV). {\it Central panel}: Radial profile of the fraction of the soft excess flux relative to the hot ICM flux in the $0.3$--$1.0$ keV band. The soft excess is modelled as thermal emission. {\it Right panel}:  The radial distribution of the deprojected electron density of the warm gas, assuming the soft excess is of thermal origin and assuming spherical symmetry of its distribution. } 
\label{proj_fluxes}
\end{figure*}

We determine the radial profile of the soft component using XMM-Newton EPIC spectra extracted from circular annuli, centred on the core of the cluster. We use annuli with outer radii of 0.5\arcmin, 1\arcmin, 2\arcmin, 3\arcmin, 4\arcmin\ and 6\arcmin. The cluster emission in each annulus is modelled with the multi-temperature {\it{wdem}} model (see Sect. \ref{spex}). 

To account for projection effects, we deproject our spectra under the assumption of spherical symmetry. 
First, we fit the spectrum in the outermost (6th) annulus and determine the normalisation, temperature and iron abundance of the hot cluster gas and the normalisation of the soft excess component. We determine which fraction of this emission is projected in front and behind of the 5th annulus. We include the projected emission of the 6th shell, with fixed parameters, in the model for the 5th annulus. In the same way, in the model for each annulus, we include the projected hot ICM and soft excess emission from all intervening shells. For the soft excess component, we try to fit both a power-law and a thermal plasma model. For the temperature of the thermal model and for the differential photon index of the non-thermal model, we adopt the best fit values obtained by fitting the spectrum extracted from the $3$\arcmin--$8$\arcmin\ region around the cluster 
(k$T=0.19$~keV and $\Gamma=2.41$), where the relative contribution of the soft excess emission is high and allows for the best determination of its spectral properties.

On the left panel of Fig.~\ref{proj_fluxes} we show the radial profile of the projected soft excess and hot ICM flux in the $0.3$--$1.0$ keV band. We model the soft excess emission with a thermal model. 
On the central panel we show the radial profile of the fraction of the soft excess flux relative to the ICM flux in the $0.3$--$1.0$ keV band. Fig.~\ref{proj_fluxes} shows that while the soft excess emission peaks at the core of the cluster, the soft excess flux relative to the cluster emission is increasing with radius beyond $\sim$3\arcmin. 

\begin{figure*}[tb]
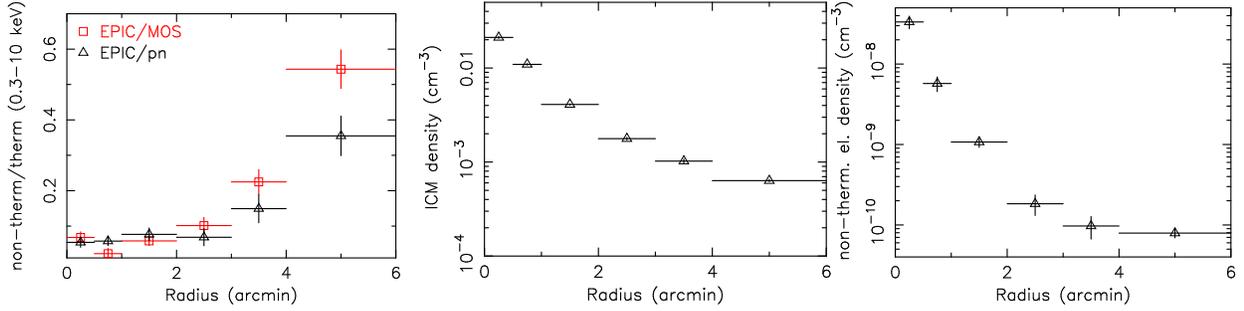

\begin{center}
\begin{minipage}{0.3\textwidth}
\includegraphics[width=0.75\textwidth,clip=t,angle=270.]{fractionnontherm_therm.ps}
\end{minipage}
\begin{minipage}{0.3\textwidth}
\includegraphics[width=0.75\textwidth,clip=t,angle=270.]{deprojectedthermaldensity.ps}
\end{minipage}
\begin{minipage}{0.3\textwidth}
\includegraphics[width=0.75\textwidth,clip=t,angle=270.]{deprojectedNONthermaldensity.ps}
\end{minipage}
\end{center}
\caption{{\it Left panel}:Radial profile of the relative fraction of non-thermal emission in the $0.3$--$10$ keV band, if the soft excess is modelled as non-thermal emission. In order to illustrate the systematic uncertainties, we plot the EPIC/MOS and EPIC/pn data separately. {\it Central panel}: The radial distribution of the deprojected electron density of the hot ICM plasma. The soft excess was fitted with a power-law model.   {\it Right panel}: The radial distribution of the deprojected density of the relativistic non-thermal electrons.} 
\label{densities}
\end{figure*}

\begin{figure*}[tb]
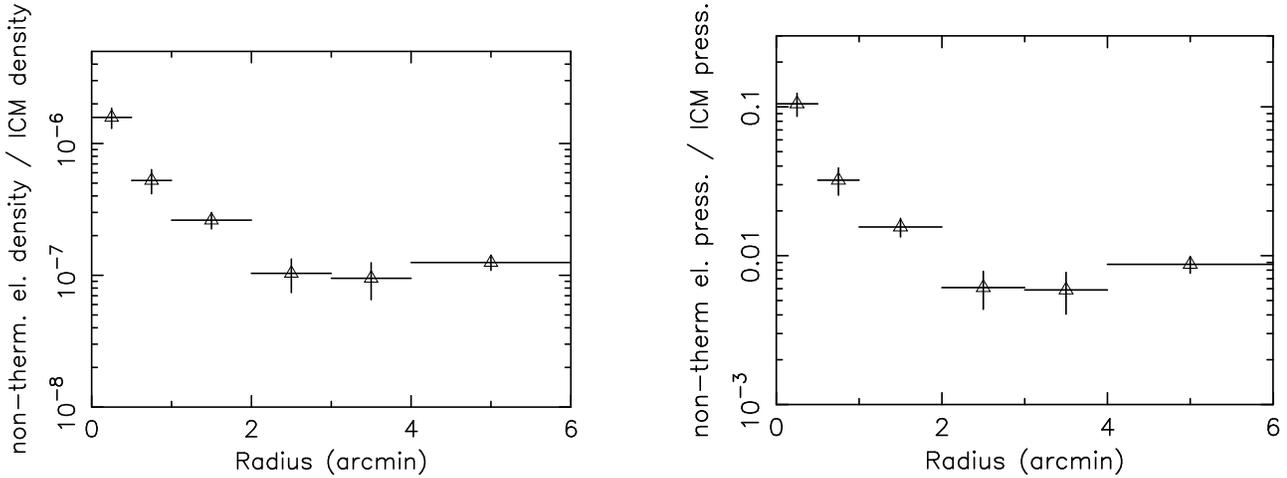

\begin{minipage}{0.5\textwidth}
\includegraphics[width=0.7\textwidth,clip=t,angle=270.]{densities.ps}
\end{minipage}
\begin{minipage}{0.5\textwidth}
\includegraphics[width=0.7\textwidth,clip=t,angle=270.]{pressures.ps}
\end{minipage}\\
\caption{{\it Left panel}: Radial profile of the ratio of the density of the relativistic electrons and of the hot ICM plasma. {\it Right panel}: Radial profile of the ratio of the pressure of the relativistic electrons and the hot ICM plasma.} 
\label{denspres}
\end{figure*}

Using the thermal description of the soft excess emission and assuming that the warm gas resides in the same shells as the hot ICM plasma, we calculate the electron density of the warm gas in each shell (see the right panel in Fig.~\ref{proj_fluxes}). Using this deprojected thermal electron density profile, we obtain a total mass for the warm gas in the central region within a radius of 2\arcmin\ (150~kpc) of $M_{\mathrm{warm}}=4.4\times10^{11}$~M$_{\odot}$. The mass of the hot ICM in the same region is $M_{\mathrm{hot}}=2.6\times10^{12}$~M$_{\odot}$.

On the left panel of Fig.~\ref{densities} we show the relative contribution of the non-thermal flux between $0.3$--$10$ keV, assuming the soft excess is of non-thermal origin. In order to illustrate the uncertainties in the radial profiles, we indicate the values obtained by EPIC/MOS and EPIC/pn separately. The plot shows that there are systematic differences in the best fit flux of the soft excess component determined using EPIC/MOS and EPIC/pn, however, both instruments show the same radial trends. The best fit values of the soft excess in the radial bins have a systematic uncertainty of the order of $\sim$40\%. The secure determination of the soft excess flux depends critically on the instrument calibration.
 
Using the non-thermal description for the soft excess emission, we calculate the average density of the hot ICM (see the central panel in Fig.~\ref{densities}) and the luminosity of the power-law component in each of the investigated shells.
We use the deprojected luminosities of the power-law component to determine the energy in the population of relativistic electrons for each of the shells \citep{lieu1999}:
\begin{equation}
\label{relelenergy}
E_{e}=8\times10^{61}L_{42}\left( \frac{3-\mu}{2-\mu} \right)\frac{\gamma^{2-\mu}_{\mathrm{max}}-\gamma^{2-\mu}_{\mathrm{min}}}{\gamma^{3-\mu}_{\mathrm{max}}-\gamma^{3-\mu}_{\mathrm{min}}}~\mathrm{ergs,}
\end{equation}
where $L_{42}$ is the luminosity of the non-thermal radiation in units of $10^{42}$ ergs~s$^{-1}$, $\gamma_{\mathrm{min}}$ and $\gamma_{\mathrm{max}}$ are the Lorentz factors corresponding to the lower and higher limits of the energy range of XMM-Newton ($0.3$--$10$ keV) and $\mu$ is the index of the differential number electron distribution. For the Lorentz factors we adopt values $\gamma_{\mathrm{min}}=600$ and $\gamma_{\mathrm{max}}=3500$, using the relation $\gamma=300(E/75\, \mathrm{eV})^{1/2}$, where $E$ is the energy to which a CMB photon is up-scattered by a relativistic electron with a Lorentz factor of $\gamma$. The relationship between the differential photon index $\Gamma$ of the observed power-law emission and the index of the electron distribution is $\mu=-1+2\Gamma$. 
We calculate the density of the relativistic electrons in the given shell by dividing the energy density in relativistic electrons by $<$$\gamma$$>$$m_{\mathrm{e}}c^2$, where $<$$\gamma$$>$ is the average Lorentz factor ($\approx1000$) and $m_{\mathrm{e}}$ is the rest mass of an electron. 

The deprojected densities of the relativistic electrons are shown on the right panel of Fig~\ref{densities}. On the left panel of Fig.~\ref{denspres}, we show the radial distribution of the ratio of the density of relativistic electrons and of the hot ICM plasma. On the right panel, we show the radial distribution of the ratio of the non-thermal to thermal electron pressure, where the pressure of the non-thermal electrons was calculated as $P=E_{e}/3V$. 

The relative density and relative pressure of the relativistic electrons peaks strongly in the core of the cluster in the area with a radius of $~2$\arcmin.

\subsection{2D maps of the soft excess emission}

\begin{figure*}[tb]
\begin{center}
\begin{minipage}{0.33\textwidth}
\includegraphics[width=0.85\textwidth,clip=t,angle=0.]{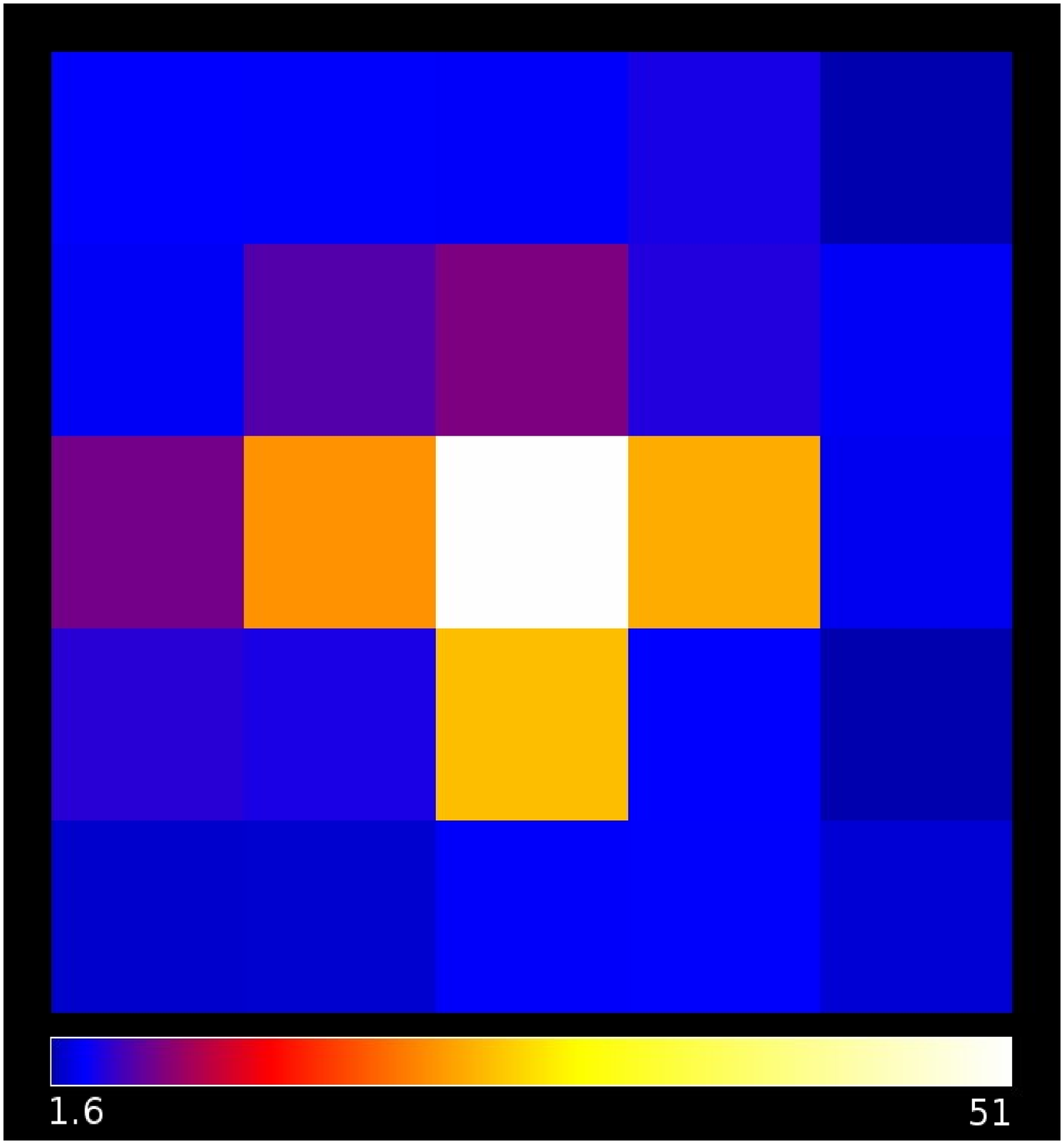}
\end{minipage}
\begin{minipage}{0.33\textwidth}
\includegraphics[width=0.85\textwidth,clip=t,angle=0.]{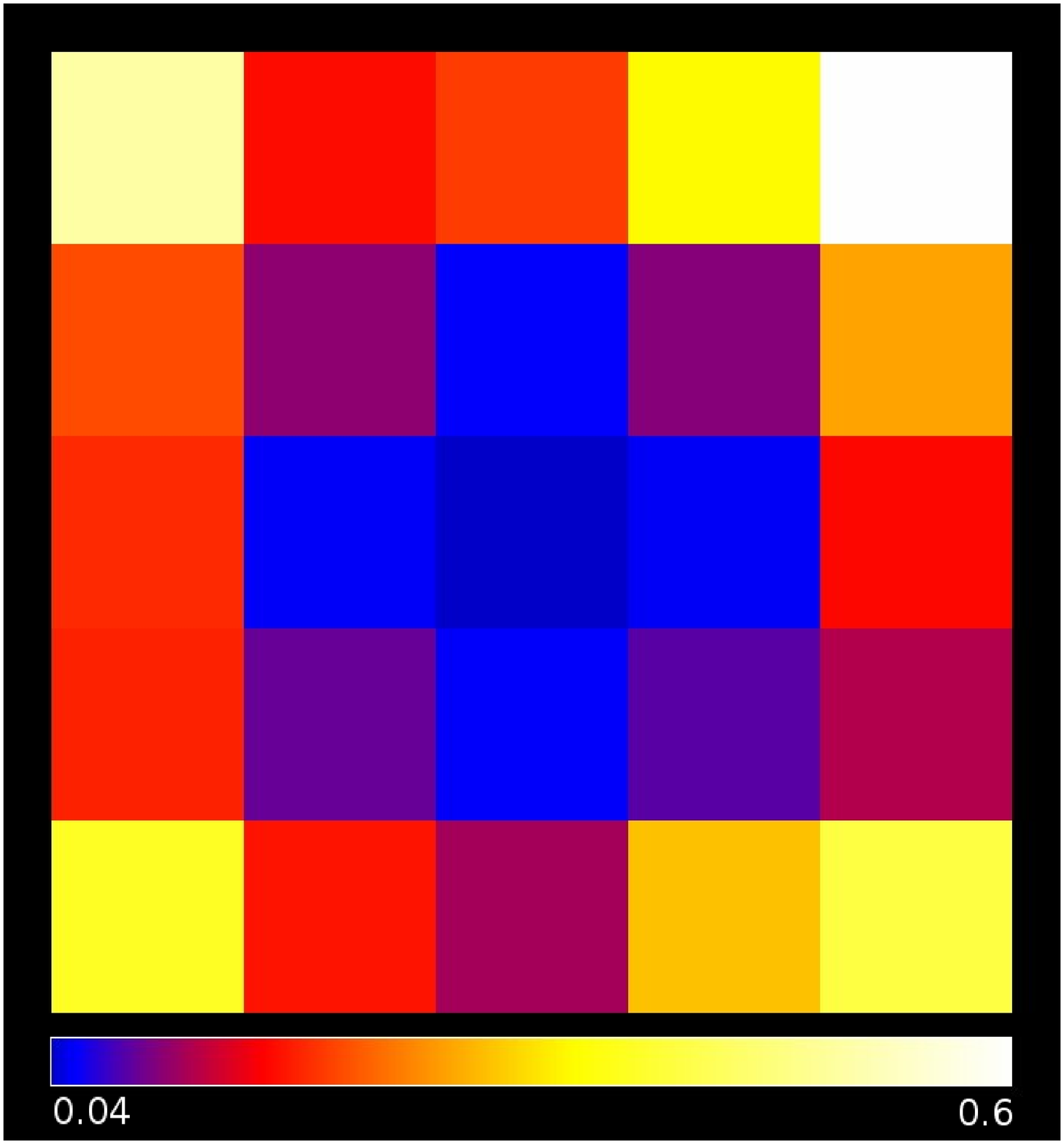}
\end{minipage}
\begin{minipage}{0.33\textwidth}
\includegraphics[width=0.95\textwidth,clip=t,angle=0.]{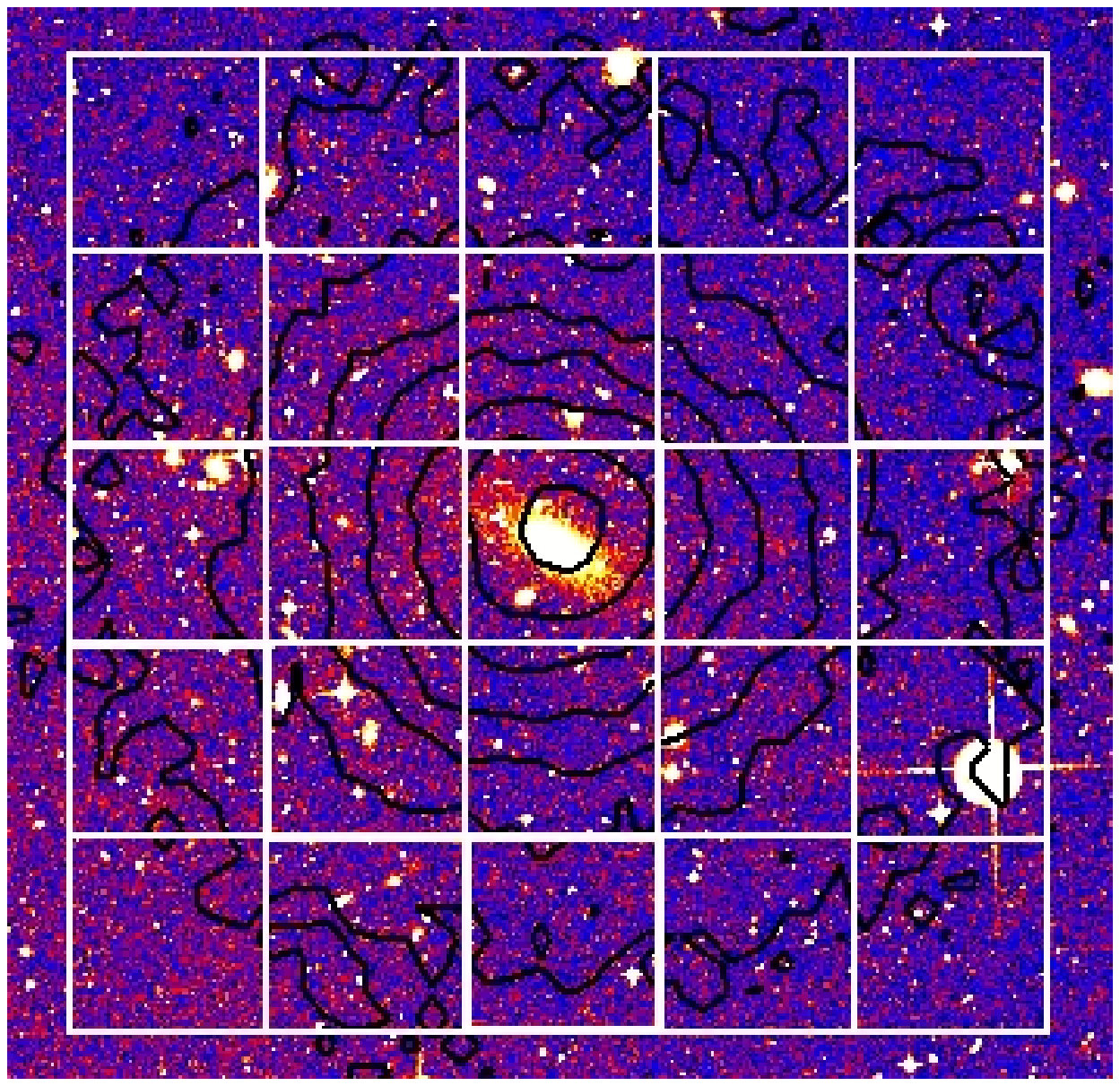}
\end{minipage}\\
\end{center}
\caption{{\it Left panel}: Map of the emission measure of the soft excess in the inner $15$\arcmin$ \times 15$\arcmin\ region of the cluster. The units are $10^{63}$ cm$^{-3}$ arcmin$^{-2}$. {\it Central panel}: Map of the ratio of the hot ICM plasma emission measure and soft excess emission measure in the same region as the previous map. {\it Right panel}: Optical Digitalised Sky Survey image of the central region of the cluster that we used to derive the map of the soft excess emission. The extraction regions for the maps and the X-ray contours from Suzaku XIS1 are overplotted. Both the soft excess and the cluster emission peak on the central cluster galaxy. }
\label{3maps}
\end{figure*}

In order to determine the 2D distribution of the soft excess emission and to look for its spatial variations, we extract 25 spectra using XMM-Newton EPIC from a rectangular grid with a bin-size of 3\arcmin\ in the 15\arcmin$\times$15\arcmin\ region centred on the core of the cluster. For each bin, we compute a spectral redistribution file and an ancillary response file. We fit the spectrum of each bin individually. We model the hot ICM plasma emission with the multi-temperature $wdem$ model and the soft excess emission with a thermal component (k$T=0.19$~keV, metallicity $Z=0.3$ solar). As shown on the left panel of Fig. \ref{3maps}, we confirm the peaked  distribution of the emission of the soft component. We also find no significant azimuthal variations in the soft excess emission. On the central panel of Fig. \ref{3maps}, we show the ratio of the emission measure of the warm gas and of the hot ICM. We again see that although the soft excess emission strongly peaks in the cluster core, its relative contribution is there the smallest and increases with the cluster radius. Both the soft excess and the cluster emission peak on the central cluster galaxy. On the right panel of Fig.~\ref{3maps}, we show an optical Digitalised Sky Survey image of the region of the cluster used to derive the map of the soft excess emission with overplotted extraction regions and X-ray contours from Suzaku XIS1.

\subsection{Search for line emission from the warm gas with RGS}
\label{RGSsubsec}

\begin{figure}
\includegraphics[width=6.7cm,clip=t,angle=270.]{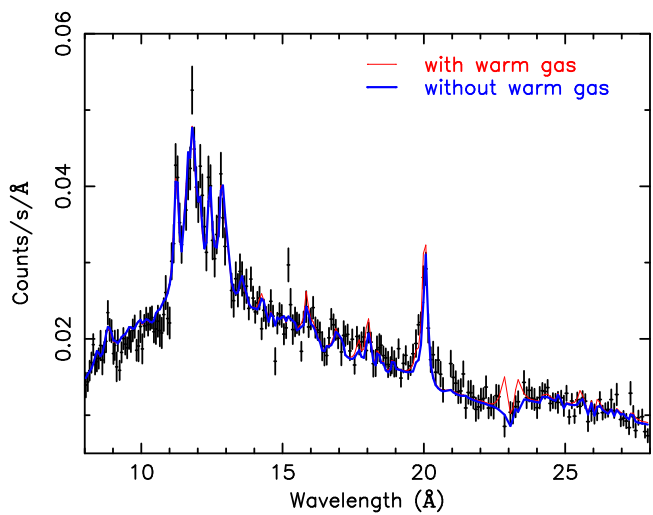}
\caption{RGS spectrum extracted from a 4\arcmin\ wide stripe centred on the cluster core. The thick and the thin lines depict the best fit models of the RGS data without and with the warm gas component, respectively. Note the mismatch between the two models at $\sim23$~\AA, the energy of the redshifted \ion{O}{vii} emission.} 
\label{RGS}
\end{figure}

Both the radial profiles and the 2D maps of the soft excess emission show that the emission is strongly peaked on the cluster core. If the soft excess emission is due to cooling gas in the core of the cluster, we could detect its presence via redshifted \ion{O}{vii} lines with the RGS detector on XMM-Newton. 
We fit the RGS spectrum extracted from a 4\arcmin\ wide stripe centred on the cluster core with the {\it{wdem}} model \citep[for the details of the RGS spectroscopy of S\'ersic~159-03 see][]{deplaa2006}. We obtain a good fit with parameters consistent with those reported by \citet{deplaa2006}. The RGS spectrum is shown in Fig.~\ref{RGS}, with the best fit {\it{wdem}} model depicted by the thick line. By fitting a Gaussian at the expected wavelength of the \ion{O}{vii} intercombination line at the cluster redshift, we find a $2\sigma$ upper limit for the \ion{O}{vii} line flux of $1.4\times10^{-6}$~counts~s$^{-1}$~cm$^{-2}$~arcmin$^{-2}$ (assuming \ion{O}{vii} emission from a circular region with a radius of 2\arcmin).

Using the sum of the deprojected emission measures of the warm gas obtained for the 3 extraction regions within the inner 2\arcmin\ of the cluster ($Y_{\mathrm{exc}\, 2^{\prime}}=3.7\times10^{65}$~cm$^{-3}$), we model the thermal spectrum of the soft excess emission. In the model we assume the parameters (k$T=0.19$, $Z=0.3$) used in Sect.~\ref{prof} to determine the deprojected radial distribution profiles. The thin line in Fig.~\ref{RGS} shows the best fit model to the RGS spectrum with the warm gas component included in the model. The model for the warm gas predicts an \ion{O}{vii} line flux of $3.2\times10^{-6}$~counts~s$^{-1}$~cm$^{-2}$~arcmin$^{-2}$. This line flux, and thus the presence of warm gas with k$T=0.2$~keV, metallicity of 0.3 solar, and mass of $M_{\mathrm{warm}}=4.4\times10^{11}$~M$_{\odot}$, can be ruled out using the RGS at the $4.6\sigma$ confidence level. 

\section{Discussion}
\label{discussion}

\subsection{Presence of the soft excess}

Although the systematic uncertainties in the determined soft excess flux are large (about 30\%), the Suzaku observation confirms the presence of the soft excess emission in S\'ersic~159-03 and the flux derived using Suzaku data is within the systematic uncertainties consistent with that found with XMM-Newton. The presence of the soft excess in this cluster is now established by Suzaku, by two independent observations with XMM-Newton and by a previous detection with ROSAT. The flux of the soft excess emission in the $9$\arcmin--$12$\arcmin\ radius is at least as large as the total flux of the Galactic foreground emission (it is more than 5 times larger than the flux of the hard distant foreground component) and the soft excess strongly peaks on the cluster core. The observed high flux and the radial distribution of the excess emission rule out the possibility that it is due to a wrong subtraction of Galactic foreground emission, or due to charge exchange emission as previously suggested \citep{bregman2006}. We conclude that the soft excess emission is present and is intrinsic to the cluster of galaxies S\'ersic~159-03.

\subsection{Thermal or non-thermal origin?}

\subsubsection{Thermal models}

We find that the soft excess emission peaks strongly on the central cD galaxy of S\'ersic~159-03. If the observed soft excess would be due to warm-hot intergalactic medium (WHIM) intercluster filaments, then the densest central part of the filament would have to be exactly along our line of sight to the cluster core and as shown by \citet{bonamente2005}, for an assumed density of $n=10^{-4}$~cm$^{-3}$ in its centre, the filament would be several hundred megaparsecs long. For smaller assumed densities it would be even longer than 1 Gpc. This scenario is not consistent with current cosmological models and observations. By comparing the observed properties of the soft excess emission reported in the literature with the predictions based on models of the WHIM by \citet{cen1999}, \citet{mittaz2004} concluded that the emission of the WHIM in current models is 3--4 orders of magnitude too faint to explain the soft excess. 

The preferred interpretation of \citet{bonamente2005} for the soft excess is warm gas with a density of $10^{-3}$--$10^{-2}$~cm$^{-3}$ in a clumpy intracluster medium with a volume filling factor much smaller than 1 as suggested by \citet{cheng2005}. According to this interpretation the warm gas would have a cooling time shorter than the Hubble time and the soft excess would be a dynamical phenomenon observable due to recent infall of smaller groups. However, only a model where the subgroup is falling in just along our line of sight would be consistent with the centrally peaked distribution of the soft emission, without significant azimuthal variations. But in that case, we would not observe the soft excess out to large radii as we do. This model is not consistent with the observations. 

From observations, the thermal interpretation can only be confirmed by the presence of \ion{O}{vii} line emission at the cluster redshift. However, the Suzaku observation does not confirm the presence of these lines. The XMM-Newton RGS data rule out the presence of the large amount of warm gas in the cluster core deduced from the deprojected radial profiles assuming $Z=0.3$ and k$T=0.19$ at the $4.6\sigma$ confidence level. Using the RGS, we found an upper limit for the \ion{O}{vii} line flux in the cluster core of $3.2\times10^{-6}$~counts~s$^{-1}$~cm$^{-2}$~arcmin$^{-2}$. 

\subsubsection{Non-thermal model}

The data are consistent with the non-thermal description of the soft excess. 
If the soft excess is of non-thermal origin, the total energy in the relatitivistic electrons producing X-ray emission between 0.3 and 10.0 keV within 8\arcmin\ (600~kpc), calculated using equation (\ref{relelenergy}), and using the luminosities and photon indeces determined in Sect.~\ref{xis1spectr}, is $\sim$$2\times10^{60}$~ergs. If we assume that the relativistic electrons have the same differential electron distribution down to energies with Lorentz factors of $\gamma\sim200$, below which the electrons loose energy rapidly due to Coulomb losses, the total energy in relativistic electrons will still not exceed $1\times10^{61}$~ergs. The total thermal energy within the radius of 
600~kpc of the cluster is $3\times10^{63}$~ergs, which means that even if the energy in the relativistic ions is as much as $\sim$30 times higher than that in relativistic electrons, the total energy in cosmic ray particles will only account for 10\% of the thermal energy of the ICM.
The distribution of the relativistic electrons peaks on the central cluster galaxy. Outside the region with a radius of 2\arcmin\ (150~kpc), the density of non-thermal electrons relative to the electron density of the hot ICM stays constant. The relative density and pressure of the relativistic electrons at the central cluster galaxy are $\approx10$ times higher than in the outer parts. This result is in agreement with that of \citet{bonamente2005}. 

The high non-thermal electron density in the cluster core might have been provided by an AGN in the central cD galaxy. Our best fit power-law photon indices in the inner and outer extraction region of $\Gamma=2.11$ and $\Gamma=2.41$, respectively, correspond to differential electron number distributions of $\mu=3.22$ and $\mu=3.82$, respectively. These are steeper than the distribution of the Galactic cosmic-ray electrons ($\mu\sim2.7$). The steepening of the power-law distribution might indicate that the relativistic electrons suffered radiative losses \citep{sarazin1999}. Relativistic electrons emitting in the observed energy range via IC emission have relatively long lifetimes. The respective inverse compton $t_{\mathrm{IC}}$ and synchrotron lifetimes are: $t_{\mathrm{IC}}=2.3\times10^{9}\,(\gamma/10^3)^{-1}\, (1+z)^{-4}$~yr and $t_{\mathrm{syn}}=2.4\times10^{10}\, (\gamma/10^3)^{-1}\, (B/1\mu\mathrm{G})^{-2}$~yr \citep[e.g.~][]{sarazin1999}. 

The relativistic particles could be seeded by jets of active galaxies, then transported outwards whilst undergoing in-situ second order Fermi acceleration by turbulent Alfven waves \citep{lieu2007}.
The presence of relativistic particles at large distances from the cluster core might also be explained by shock acceleration. For example by accretion shocks or as suggested by \citet{deplaa2006}, by shocks associated with ram pressure stripping of infalling galaxies. 
Diffuse shocks accelerate both electrons and ions and the energy in ions could be $\sim$$20$--$50$ times larger than the energy in electrons \citep{bell1978b}. If we take this into account, then the pressure in the cluster center will be close to equipartition with the hot ICM pressure.
However, if a significant fraction of the relativistic electrons associated with the central peak is originating in AGN electron/positron jets then the pressure in the relativistic particles might still be below pressure equipartition in the cluster core.

Recently, \citet{bagchi2006} reported that the ring-shaped non-thermal radio emitting structure found at the outskirts of the rich cluster of galaxies Abell~3376 may trace the shock waves of cosmological large scale matter flows. The giant radio structures in Abell~3376 are probably due to merger or accretion shocks near the virial infall region of the cluster, which is the transition zone between the hot cluster medium and the WHIM. Radio sources, like that in Abell~3376 may also be the sites where magnetic shocks accelerate cosmic ray particles \citep{bagchi2006}. The example of Abell~3376 and of other clusters with radio halos shows that in clusters of galaxies electrons are accelerated to relativistic energies with a Lorentz factors of the order of $10^{4}$ even at large radii. The strong soft excess in the outer parts of S\'ersic~159-03 might be due to accretion shocks. In order to get a more complete picture about the physical processes in S\'ersic~159-03, it is important to obtain radio data of the cluster, that will probe the population of the relativistic electrons with higher energies than those the presence of which we infer from the detected soft excess emission. 

Clusters of galaxies are more complicated than previously thought. 
If the soft excess emission is of non-thermal origin, as our data suggest, then the non-thermal component needs to be taken into account in the overall mass and energy budget of galaxy clusters. Reduction of the emission measure of the hot cluster gas in the outer parts of the cluster by 40\% means that the amount of gas in the outskirts is 20\% less than indicated by models which do not take into account the presence of the non-thermal emission. The cluster masses are calculated by assuming hydrostatic equilibrium, not taking into account the possible non-thermal pressure. The non-thermal pressure may be high enough to affect the estimates of the total cluster mass.   

If a substantial fraction of the central $2$--$10$ keV emission of some clusters of galaxies
is of non-thermal origin, it will not only account for the detection of soft and hard excesses, but also for the significantly less than expected Sunyaev-Zel'dovich effect, that emerged
from a comparison of WMAP and X-ray data for a large sample of nearby
clusters \citep{lieu2006,afshordi2007}. Because for an equal pressure of thermal and
relativistic electrons the Sunyaev-Zel'dovich decrement is much smaller due to the relativistic electrons \citep[e.g.][]{ensslin2000}.

\section{Conclusions}

We have analysed new Suzaku XIS1 spectra and archival XMM-Newton EPIC and RGS spectra of the cluster of galaxies S\'ersic 159-03. We found that:

\begin{itemize}
\item
The Suzaku observation confirms the presence of the soft excess emission and its derived flux is consistent with the values determined using XMM-Newton. Suzaku does not confirm the presence of the redshifted \ion{O}{vii} lines in the cluster.
 
\item 
Radial profiles and 2D maps show that the soft excess emission has a strong peak at the position of the central cD galaxy and the maps do not show any significant azimuthal variations. Although the soft excess emission can be fitted equally well with both thermal and non-thermal models, the spatial distribution of the soft emission is not consistent with any of the viable thermal models: intercluster WHIM filaments, or models of clumpy warm intracluster gas associated with infalling groups. 

\item
Using the XMM-Newton RGS data we do not confirm the presence of the warm gas in the cluster centre as inferred from the deprojected EPIC density profiles and we put upper limits on the \ion{O}{vii} line emission from the cluster core. 

\item
The observed properties of the soft excess emission are consistent with the non-thermal interpretation. While the high density of relativistic electrons associated with the peak of the soft emission in the cluster centre might have been provided by an AGN in the central cD galaxy, the underlying population might have been accelerated in diffuse shocks. 

\end{itemize}

\begin{acknowledgements}
We would like to thank the anonymous referee for providing comments that helped to improve the clarity of the paper.
This work is based on observations obtained with XMM-Newton, an ESA science mission with instruments
and contributions directly funded by ESA member states and the USA (NASA). The Netherlands Institute
for Space Research (SRON) is supported financially by NWO, the Netherlands Organization for Scientific Research. 
\end{acknowledgements}

\bibliographystyle{aa}
\bibliography{clusters}

\begin{thebibliography}{45}
\expandafter\ifx\csname natexlab\endcsname\relax\def\natexlab#1{#1}\fi

\bibitem[{{Afshordi} {et~al.}(2006){Afshordi}, {Lin}, {Nagai}, \&
  {Sanderson}}]{afshordi2007}
{Afshordi}, N., {Lin}, Y.-T., {Nagai}, D., \& {Sanderson}, A.~J.~R. 2006,
  MNRAS, submitted, astro-ph/0612700

\bibitem[{{Bagchi} {et~al.}(2006){Bagchi}, {Durret}, {Neto}, \&
  {Paul}}]{bagchi2006}
{Bagchi}, J., {Durret}, F., {Neto}, G.~B.~L., \& {Paul}, S. 2006, Science, 314,
  791

\bibitem[{{Bell}(1978)}]{bell1978b}
{Bell}, A.~R. 1978, \mnras, 182, 443

\bibitem[{{Bonamente} {et~al.}(2001){Bonamente}, {Lieu}, \&
  {Mittaz}}]{bonamente2001}
{Bonamente}, M., {Lieu}, R., \& {Mittaz}, J.~P.~D. 2001, \apjl, 561, L63

\bibitem[{{Bonamente} {et~al.}(2005){Bonamente}, {Lieu}, {Mittaz}, {Kaastra},
  \& {Nevalainen}}]{bonamente2005}
{Bonamente}, M., {Lieu}, R., {Mittaz}, J.~P.~D., {Kaastra}, J.~S., \&
  {Nevalainen}, J. 2005, \apj, 629, 192

\bibitem[{{Bregman} \& {Lloyd-Davies}(2006)}]{bregman2006}
{Bregman}, J.~N. \& {Lloyd-Davies}, E.~J. 2006, \apj, 644, 167

\bibitem[{{Cen} \& {Ostriker}(1999)}]{cen1999}
{Cen}, R. \& {Ostriker}, J.~P. 1999, \apj, 514, 1

\bibitem[{{Cheng} {et~al.}(2005){Cheng}, {Borgani}, {Tozzi}, {Tornatore},
  {Diaferio}, {Dolag}, {He}, {Moscardini}, {Murante}, \& {Tormen}}]{cheng2005}
{Cheng}, L.-M., {Borgani}, S., {Tozzi}, P., {et~al.} 2005, \aap, 431, 405

\bibitem[{{de Grandi} {et~al.}(1999){de Grandi}, {B{\"o}hringer}, {Guzzo},
  {Molendi}, {Chincarini}, {Collins}, {Cruddace}, {Neumann}, {Schindler},
  {Schuecker}, \& {Voges}}]{degrandi1999}
{de Grandi}, S., {B{\"o}hringer}, H., {Guzzo}, L., {et~al.} 1999, \apj, 514,
  148

\bibitem[{{De Luca} \& {Molendi}(2004)}]{deluca2004}
{De Luca}, A. \& {Molendi}, S. 2004, \aap, 419, 837

\bibitem[{{de Plaa} {et~al.}(2005){de Plaa}, {Kaastra}, {M{\'e}ndez}, {Tamura},
  {Bleeker}, {Peterson}, {Paerels}, {Bonamente}, \& {Lieu}}]{deplaa2005}
{de Plaa}, J., {Kaastra}, J.~S., {M{\'e}ndez}, M., {et~al.} 2005, Adv. Space
  Res., 36, 601

\bibitem[{{de Plaa} {et~al.}(2006){de Plaa}, {Werner}, {Bykov}, {Kaastra},
  {M{\'e}ndez}, {Vink}, {Bleeker}, {Bonamente}, \& {Peterson}}]{deplaa2006}
{de Plaa}, J., {Werner}, N., {Bykov}, A.~M., {et~al.} 2006, \aap, 452, 397

\bibitem[{{Dickey} \& {Lockman}(1990)}]{dickey1990}
{Dickey}, J.~M. \& {Lockman}, F.~J. 1990, \araa, 28, 215

\bibitem[{{Ensslin} \& {Biermann}(1998)}]{ensslin1998}
{Ensslin}, T.~A. \& {Biermann}, P.~L. 1998, \aap, 330, 90

\bibitem[{{En{\ss}lin} \& {Kaiser}(2000)}]{ensslin2000}
{En{\ss}lin}, T.~A. \& {Kaiser}, C.~R. 2000, \aap, 360, 417

\bibitem[{{Fujimoto} {et~al.}(2007){Fujimoto}, {Mitsuda}, {McCammon}, {Takei},
  {Bauer}, {Ishisaki}, {Porter}, {Yamaguchi}, {Hayashida}, \&
  {Yamasaki}}]{fujimoto2006}
{Fujimoto}, R., {Mitsuda}, K., {McCammon}, D., {et~al.} 2007, \pasj, 59, 133

\bibitem[{{Hwang}(1997)}]{hwang1997}
{Hwang}, C.-Y. 1997, Science, 278, 1917

\bibitem[{{Ishisaki} {et~al.}(2007){Ishisaki}, {Maeda}, {Fujimoto}, {Ozaki},
  {Ebisawa}, {Takahashi}, {Ueda}, {Ogasaka}, {Ptak}, {Mukai}, {Hamaguchi},
  {Hirayama}, {Kotani}, {Kubo}, {Shibata}, {Ebara}, {Furuzawa}, {Iizuka},
  {Inoue}, {Mori}, {Okada}, {Yokoyama}, {Matsumoto}, {Nakajima}, {Yamaguchi},
  {Anabuki}, {Tawa}, {Nagai}, {Katsuda}, {Hayashida}, {Bamba}, {Miller},
  {Sato}, \& {Yamasaki}}]{ishisaki2006}
{Ishisaki}, Y., {Maeda}, Y., {Fujimoto}, R., {et~al.} 2007, \pasj, 59, 113

\bibitem[{{Jansen} {et~al.}(2001){Jansen}, {Lumb}, {Altieri}, {Clavel}, {Ehle},
  {Erd}, {Gabriel}, {Guainazzi}, {Gondoin}, {Much}, {Munoz}, {Santos},
  {Schartel}, {Texier}, \& {Vacanti}}]{jansen2001}
{Jansen}, F., {Lumb}, D., {Altieri}, B., {et~al.} 2001, \aap, 365, L1

\bibitem[{{Kaastra} {et~al.}(2001){Kaastra}, {Ferrigno}, {Tamura}, {Paerels},
  {Peterson}, \& {Mittaz}}]{kaastra2001}
{Kaastra}, J.~S., {Ferrigno}, C., {Tamura}, T., {et~al.} 2001, \aap, 365, L99

\bibitem[{{Kaastra} {et~al.}(2003){Kaastra}, {Lieu}, {Tamura}, {Paerels}, \&
  {den Herder}}]{kaastra2003}
{Kaastra}, J.~S., {Lieu}, R., {Tamura}, T., {Paerels}, F.~B.~S., \& {den
  Herder}, J.~W. 2003, \aap, 397, 445

\bibitem[{{Kaastra} {et~al.}(1996){Kaastra}, {Mewe}, \&
  {Nieuwenhuijzen}}]{kaastra1996}
{Kaastra}, J.~S., {Mewe}, R., \& {Nieuwenhuijzen}, H. 1996, in UV and X-ray
  Spectroscopy of Astrophysical and Laboratory Plasmas p.411, K. Yamashita and
  T. Watanabe. Tokyo : Universal Academy Press

\bibitem[{{Kaastra} {et~al.}(2004){Kaastra}, {Tamura}, {Peterson}, {Bleeker},
  {Ferrigno}, {Kahn}, {Paerels}, {Piffaretti}, {Branduardi-Raymont}, \&
  {B\"ohringer}}]{kaastra2004}
{Kaastra}, J.~S., {Tamura}, T., {Peterson}, J.~R., {et~al.} 2004, \aap, 413,
  415

\bibitem[{{Koyama} {et~al.}(2007){Koyama}, {Tsunemi}, {Dotani}, {Bautz},
  {Hayashida}, {Tsuru}, H., {Ogawara}, \& {et al.}}]{koyama2007}
{Koyama}, K., {Tsunemi}, H., {Dotani}, T., {et~al.} 2007, \pasj, 59, 23

\bibitem[{{Kuntz} \& {Snowden}(2000)}]{kuntz2000}
{Kuntz}, K.~D. \& {Snowden}, S.~L. 2000, \apj, 543, 195

\bibitem[{{Lieu} {et~al.}(1999){Lieu}, {Ip}, {Axford}, \&
  {Bonamente}}]{lieu1999}
{Lieu}, R., {Ip}, W.-H., {Axford}, W.~I., \& {Bonamente}, M. 1999, \apjl, 510,
  L25

\bibitem[{{Lieu} {et~al.}(1996{\natexlab{a}}){Lieu}, {Mittaz}, {Bowyer},
  {Breen}, {Lockman}, {Murphy}, \& {Hwang}}]{lieu1996a}
{Lieu}, R., {Mittaz}, J.~P.~D., {Bowyer}, S., {et~al.} 1996{\natexlab{a}},
  Science, 274, 1335

\bibitem[{{Lieu} {et~al.}(1996{\natexlab{b}}){Lieu}, {Mittaz}, {Bowyer},
  {Lockman}, {Hwang}, \& {Schmitt}}]{lieu1996b}
{Lieu}, R., {Mittaz}, J.~P.~D., {Bowyer}, S., {et~al.} 1996{\natexlab{b}},
  \apjl, 458, L5

\bibitem[{{Lieu} {et~al.}(2006){Lieu}, {Mittaz}, \& {Zhang}}]{lieu2006}
{Lieu}, R., {Mittaz}, J.~P.~D., \& {Zhang}, S.-N. 2006, \apj, 648, 176

\bibitem[{{Lieu} \& {Quenby}(2006)}]{lieu2007}
{Lieu}, R. \& {Quenby}, J. 2006, ApJ, submitted, astro-ph/0607304

\bibitem[{{Lockman} \& {Savage}(1995)}]{lockman1995}
{Lockman}, F.~J. \& {Savage}, B.~D. 1995, \apjs, 97, 1

\bibitem[{{Lodders}(2003)}]{lodders2003}
{Lodders}, K. 2003, \apj, 591, 1220

\bibitem[{{Lumb} {et~al.}(2002){Lumb}, {Warwick}, {Page}, \& {De
  Luca}}]{lumb2002}
{Lumb}, D.~H., {Warwick}, R.~S., {Page}, M., \& {De Luca}, A. 2002, \aap, 389,
  93

\bibitem[{{Maia} {et~al.}(1987){Maia}, {da Costa}, {Willmer}, {Pellegrini}, \&
  {Rite}}]{maia1987}
{Maia}, M.~A.~G., {da Costa}, L.~N., {Willmer}, C., {Pellegrini}, P.~S., \&
  {Rite}, C. 1987, \aj, 93, 546

\bibitem[{{Mitsuda} {et~al.}(2007){Mitsuda}, {Bautz}, {Inoue}, {Kelley},
  {Koyama}, \& {Kunieda}}]{mitsuda2007}
{Mitsuda}, K., {Bautz}, M., {Inoue}, H., {et~al.} 2007, \pasj, 59, 1

\bibitem[{{Mittaz} {et~al.}(2004){Mittaz}, {Lieu}, {Cen}, \&
  {Bonamente}}]{mittaz2004}
{Mittaz}, J., {Lieu}, R., {Cen}, R., \& {Bonamente}, M. 2004, \apj, 617, 860

\bibitem[{{Mittaz} {et~al.}(1998){Mittaz}, {Lieu}, \& {Lockman}}]{mittaz1998}
{Mittaz}, J.~P.~D., {Lieu}, R., \& {Lockman}, F.~J. 1998, \apjl, 498, L17

\bibitem[{{Nevalainen} {et~al.}(2006){Nevalainen}, {Bonamente}, \&
  {Kaastra}}]{nevalainen2006}
{Nevalainen}, J., {Bonamente}, M., \& {Kaastra}, J. 2006, ApJ, in press,
  astro-ph/0610461

\bibitem[{{S{\' e}rsic}(1974)}]{sersic1974}
{S{\' e}rsic}, J.~L. 1974, \apss, 28, 365

\bibitem[{{Sarazin}(1999)}]{sarazin1999}
{Sarazin}, C.~L. 1999, \apj, 520, 529

\bibitem[{{Sarazin} \& {Lieu}(1998)}]{sarazin1998}
{Sarazin}, C.~L. \& {Lieu}, R. 1998, \apjl, 494, L177

\bibitem[{{Serlemitsos} {et~al.}(2007){Serlemitsos}, {Soong}, {Chan},
  {Okajima}, {Lehan}, {Maeda}, {Itoh}, {Mori}, {Iizuka}, {Itoh}, {Inoue},
  {Okada}, {Yokoyama}, {Itoh}, {Ebara}, {Nakamura}, {Suzuki}, {Ishida},
  {Hayakawa}, {Inoue}, {Okuma}, {Kubota}, {Suzuki}, {Osawa}, {Yamashita},
  {Kunieda}, {Tawara}, {Ogasaka}, {Furuzawa}, {Tamura}, {Shibata}, {Haba},
  {Naitou}, \& {Misaki}}]{serlemitsos2007}
{Serlemitsos}, P.~J., {Soong}, Y., {Chan}, K.-W., {et~al.} 2007, \pasj, 59, 9

\bibitem[{{Takei} {et~al.}(2007){Takei}, {Ohashi}, {Henry}, {Mitsuda},
  {Fujimoto}, {Tamura}, {Yamasaki}, {Hayashida}, {Tawa}, {Matsushita}, {Bautz},
  {Hughes}, {Madejski}, {Kelley}, \& {Arnaud}}]{takei2006}
{Takei}, Y., {Ohashi}, T., {Henry}, J.~P., {et~al.} 2007, \pasj, 59, 339

\bibitem[{{Verner} {et~al.}(1996){Verner}, {Verner}, \& {Ferland}}]{verner1996}
{Verner}, D.~A., {Verner}, E.~M., \& {Ferland}, G.~J. 1996, Atomic Data and
  Nuclear Data Tables, 64, 1

\bibitem[{{Werner} {et~al.}(2006){Werner}, {de Plaa}, {Kaastra}, {Vink},
  {Bleeker}, {Tamura}, {Peterson}, \& {Verbunt}}]{werner2006}
{Werner}, N., {de Plaa}, J., {Kaastra}, J.~S., {et~al.} 2006, \aap, 449, 475

\end{thebibliography}

\appendix

\section{The contamination layer for Suzaku XIS1 at the time of the observation 
\label{contamination}}

We analysed the data obtained during a simultaneous XMM-Newton EPIC/pn, Chandra LETGS/HRC and Suzaku XIS1 observation of the blazar PKS~2155-304. The observation was performed on May 1--2, 2006, only 3 days after our observation of the cluster of galaxies S\'ersic~159-03. This observation allows us to determine the thickness of the contaminating layer in the centre of the field of view of Suzaku XIS1 at the time of our observation. 
Combining the best fit central value with the radial contamination profile, that is determined by the XIS team,
we can determine the contaminating column in each analysed region at the time of our observation.

\subsection{Data reduction}

We used version 1.2 cleaned event files to extract the XIS1 spectrum. The
spectrum was extracted from a circular region with a radius of 250
detector pixels. Although the region is not fully covered in the
observation due to the 1/4 window observation mode, the loss of flux
is only a few percent and independent of the energy. Hence, we can
neglect this effect. Since PKS~2155-304 is a highly variable object, we only used the time intervals when both XMM-Newton EPIC/pn and XIS1, or Chandra LETGS and XIS1 data were available. The RMF file for XIS1 was created using
``xisrmfgen''. We used the ARF file
``ae\_xi1\_xisnom6\_20060615.arf'' (this file is distributed by the XIS
team), which does not take into account the contamination on the optical blocking filter (OBF).

The XMM-Newton data were reduced using SAS~7.0.0. 
After selecting only those time intervals when both Suzaku and XMM-Newton were observing the source, we analysed 8.6~ks of data from both satellites.
The EPIC/pn observation was performed in the small window mode. We verified that the obtained spectrum is not affected by pile-up.

Chandra LETGS/HRC data were reprocessed according to the standard procedure
of CIAO 3.3.0 with CALDB 3.2.3. After selecting only those
periods when both LETGS and XIS1 data were available, we analysed 17.8~ks of data from both satellites.  The spectra
of plus and minus orders were merged.  We created $\pm$1st--8th order
response files (RMFs and Grating ARFs). In order to take
into account the overlap of the dispersed photons of each order, we adopted the sum of $\pm$1st--8th order
response files as the response for the LETGS spectrum.

\subsection{Simultaneous Suzaku XIS1 and XMM-Newton EPIC/pn observation}

We fitted the spectra using the SPEX spectral fitting package. 
The source spectrum was fitted with a broken power-law. For the Galactic absorption model, we used the ``hot'' model of SPEX, where we set the
abundances in the absorber to the proto-solar values of Lodders (2003).
We fixed the hydrogen column density in the model to the best determined value of $N_{\mathrm{H}}=(1.36\pm0.10)\times10^{20}$~cm$^{-2}$, which was found using a dedicated \ion{H}{i} observation toward this blazar \citep{lockman1995}. We fitted the EPIC/pn spectrum in the 0.35--10 keV band.
Our best fit low energy power-law photon index is $\Gamma=2.61\pm0.01$. The
break in the spectrum is at $2.0\pm0.3$~keV and its best fit value is $\Delta\Gamma=-0.11\pm0.03$. 

We used the best fit model determined using EPIC/pn for the XIS1 data. The XIS1 spectra were fitted in the 0.35--7 keV band. In order to find the best fit absorption column on the OBF of XIS1, we included in the model free \ion{C}{i} and \ion{O}{i} absorbing columns. Their best-fit values are $N_{\mathrm{C}}=(4.11\pm0.06)\times10^{18}$~cm$^{-2}$ and $N_{\mathrm{O}}=(6.2\pm0.5)\times10^{17}$~cm$^{-2}$.

The contamination values determined using the EPIC/pn are only slightly lower
than the nominal values from the standard contamination file for September 1.  

\subsection{Simultaneous XIS1 and Chandra LETGS/HRC observation}

We fitted the LETGS and the XIS1 spectra simultaneously using the SHERPA spectral fitting package. 
The fitting range was 0.17--2.0 keV for LETGS and 0.35--7.0 keV for XIS1.
The spectrum of \object{PKS 2155-304} was modelled as a power-law convolved with
Galactic absorption (wabs model). The model was additionally convolved
with \ion{C}{i} and \ion{O}{i} absorption by the contamination on the OBF of XIS1.  
Since the most probable composition of the contaminant is C$_{24}$H$_{38}$O$_{4}$, the C/O ratio was fixed in the fitting process to 6. 
We fixed the hydrogen column density in the model to the best determined Galactic value of $N_{\mathrm{H}}=(1.36\pm0.10)\times10^{20}$~cm$^{-2}$. 
The spectra were well fitted with a photon index of
$\Gamma=2.54\pm0.01$. The best fit contaminating columns are
$N_{\mathrm{C}}=(3.78\pm0.05)\times10^{18}$~cm$^{-2}$ and
$N_{\mathrm{O}}=6.30\times10^{17}$~cm$^{-2}$. Note that the photon index and the flux
are consistent between LETGS and XIS1 if they are fitted separately. We verified that we obtain consistent values for the absorbing column also with the $N_{\mathrm{H}}$ as a free parameter. 
We also tried broken power-law model which did not improve the fit
significantly.

The contaminating columns determined using the LETGS are very similar
to the nominal values from the standard contamination file for July 1.

\end{document}